# The Effect of Salt Shock on Growth and Pigment Accumulation of *Dunaliella Salina*

Mahsa Yazdani [1], Omid Tavakoli [2+]

1 Master student in Pharmaceutical Engineering, School of Chemical Engineering, College of Engineering, University of Tehran, Tehran, Iran.

2 Assistant Professor, School of Chemical Engineering, College of Engineering, University of Tehran, Tehran, Iran.

**Abstract.** *Dunaliella Salina* is a halotolerant microalga with great pharmaceutical and industrial potential, which commonly exists in hypersaline environments. Moreover, it is the best commercial source of beta-carotene (which has high anti-oxidant properties) in comparison to other microalgae. In this study, we investigated growth and accumulations of chlorophyll a and b, beta-carotene, and carotenoid after salt shock in 1, 1.5, 2, 2.5, 3M concentrations of NaCl. The highest cell growth rate was observed in 1 M salt shock at 22-25°C with a light intensity of $2.084\ (mW.cm)^{-2}$, a light period of 12-12, and at an initial pH of about 7.1. Although the cell growth was enhanced in 1 and 1.5M, further increase in salt content harmed cell growth. The most considerable beta-carotene quantity was attained after 1M salt shock. According to the experimental observations, it was seen that the salt shock in some concentrations is one of the practical approaches to improve the accumulation of pigments.

## 1. Introduction

*Dunaliella* species belong to the phylum Chlorophyta, order Volvocales and family Polyblepharidaceae, and are unicellular, photosynthetic and motile biflagellate microalgae morphologically distinguished by the lack of a rigid cell wall [1]. Microalgaes are a source of a variety of natural products [2]. In addition to chlorophylls a and b, the members of Dunaliella contain worth carotenoid pigments such as alfa- and beta-carotene, violaxanthin, neoxanthin, zeaxanthin and lutein [3]. In fact, Carotenoids are natural pigments with much diversified structure and widely spread in nature, where they fulfill their essential biological functions. Some of Carotenoids are provitamin A which have a range of diverse biological functions and actions, especially regarding human health [4]. Superior bioavailability, antioxidant capacity and physiological effects, substantiate the commercial interest of the algal carotene over its synthetic counterpart [5, 6]. D. salina can accumulate a very high concentration (up to14% ) of cell dry weight of beta-carotene under stress conditions of high light, high salinity, high temperature and nutrient deprivation like nitrogen starvation and it is recognized as the best biological source of this carotenoid [4, 7]. Nowadays, beta-carotene is extensively used as a colorant, a food additive, an antioxidant, an anti-cancer agent, a preventative supplement against heart disease, and for cosmetic purposes [8]. The carotenoid content of the microalgal biomass is positively correlated with the market price for the latter. Therefore, the influence of the operating conditions on Dunaliella cultures on the quality of its biomass is of major concern [9]. The world market for carotenoids was estimated at $887 million in 2004 and $1 billion in 2009 and 80% of this market is attributed to carotenoids that can be produced by using the invention process [4].

---

+ Corresponding author. Tel.: +98-21-6111 2187; fax: +98-21-6649 8984.
  *E-mail address*: otavakoli@ut.ac.ir.

Recently, some data suggested that Dunaliella has an exceptional ability to remove Na+ ions in hypersaline environments by using a novel redox-driven sodium pump [10]. Severe stress conditions such as high salinity have been claimed to induce beta-carotene production in the cell; but, at the same time, the number of cells per culture volume decreases by affecting cell division [11]. Other authors commented that the salinity does not have a clear effect on beta-carotene accumulation per cell [12]. However, all existing commercial Dunaliella facilities grow the alga outdoors in open-air cultures at high salinity: extensive culture systems (Australia and China) or more intensive, paddle-wheel stirred raceway ponds (Israel and USA) [13]. It seems that the effect of salinity, like other inducing actors on beta-carotene production in *D. salina,* is strain dependent and only a few strains of *Dunaliella* have the potential to produce up to 10% beta-carotene [14]. In fact, the halotolerant Dunaliella genus comprises green microalgae that have been intensively studied for the production of beta-carotene as a valuable compound for the health food industry [11].

The objective of this research was to recognize the best concentration of salt for growing and accumulating of pigment specially beta-carotene and evaluation of changes of this two parameters after shocking. The differences in the growth and beta-carotene accumulation of *Dunaliella* strains under different stress conditions were studied in order to determine the optimal conditions for commercial algal production for providing scientific guidance for the further optimization of production in the Persian Gulf.

## 2. Materials and Methods

### 2.1. Organism and Medium

*Dunaliella Salina* strains were obtained from the Faculty of Marine Science and Technology at Persian Gulf University. This was isolated from the Persian Gulf coast. This microalga was cultured in air-lift photo bioreactor with a working volume about 8000 ml. The culture medium was Modified Johnsons [15]. The initial pH was controlled at 7.4 and the pH level was measured with a AZ8685 pH meter manufactured by AZ INSTRUMENT CORP. also the initial cell density was $81.607 \times 10^5$ mL and the salinity (NaCl concentration) was 0.5 M (29.25 g NaCl L$^{-1}$).

In order to study the effects of salt shock on the growth and pigment accumulation of this *Dunaliella* strain, we should save all condition in the best and just change concentration of salt. Light intensify is about 2.084 mW/cm$^2$ and The light source was daylight fluorescence lamps with 12 hours' light/12 hours' dark. temperature was determined based on the conditions in an actual algal cultivation and it's about 22-25 °C. All of the components of the Modified Johnsons Medium were maintained at the original concentration.

### 2.2. Biomass and Cell Number

Cell densities were checked regularly with a haematocytometer under a Microscope. Also we curved standard chart for this. The optical density (OD) of the culture solution of different concentration was determined by spectrophotometer under wavelength 630 nm and cell number of each sample was determined directly by counting under a light microscope with haemaytometer. The equation was obtained: y = 191.1x - 26.746, with R² = 0.9968, where y is cell density, and x represents OD$_{630}$ value.

About biomass or dry weight of cell we have similar way. At first, we determined the optical density of the culture solution of different concentration by using spectrophotometer under wavelength 630 nm and then, after centrifugation at 4,000 rpm for 5 min, the contents of samples were transferred to balance filter paper, washing with distilled water and after drying about 2 hours at 60 °C in the oven, we weight them. A relationship curve between OD$_{630}$ and the biomass (g.L$^{-1}$) of dried algae was obtained: y = 71.542x - 26.18, with R² = 0.9896, where y is the biomass and x represents the OD$_{630}$ value.

### 2.3. Isolation and Determination of Pigments

For pigment extraction, first of all centrifuged certain volume of algae samples at 4,000 rpm for 5 min, then discarding supernate and the precipitate was dissolved with the same volume of acetone (90%), shaking and stewing till separated. The pigment extract was moved to a new tube. Repeating this processes and sonicated them for about 10 minutes at the end until the extract turned to white. The extracted solutions were examined by UV-2100 Spectrophotometer (UNICO, US).

The chlorophyll content was calculated by using the following equations [15, 16]: about chlorophyll *a*: Chl *a* (µg.mL$^{-1}$) = 11.24($A_{662}$) - 2.04 ($A_{645}$) and about chlorophyll *b*: Chl *b* (µg mL$^{-1}$) = 20.13($A_{645}$) - 4.19($A_{662}$). Also for calculating volume of total carotenoid we have: Cx+c (µg mL$^{-1}$) = (1000 ($A_{470}$) -1.90 Chl *a* - 63.14 Chl *b*)/214.

About isolation and determination of beta-carotene, we made beta-carotene standard solutions of different concentrations with pure acetone as solvent were detected under wavelength 453 nm where was the characteristic spectrum of beta-carotene. Actually content of beta-carotene in standard solution was determined spectrophotometrically and we curved standard chart. Finally, we determined beta-carotene by using the following equations: beta-Carotene (mg.l$^{-1}$) = 37.453($A_{453}$) + 0.1846.

All trials were evaluated 3 times to guarantee the statistical significance of the measurements.

## 3. Results and discussion

### 3.1. The Effect of Salt Concentration, pH and Light Period on Growth

Cultivation was conducted in different concentrations of NaCl and the best growth was found in 0.5M. With increasing salt concentration, growth was decreased. Also we detected best initial pH which was between 9-11 and the best result of growth has been detected about pH=10 and we provided this pH with adding NaOH and Hcl (Figure 1). Light period was 24-0 and 12-12 daylight and this has been detected in different NaCl concentrations. The best one was 12-12 in 0.5M (Figure 2). When we reached at these result, cultivation was started in 0.5M at 22-25 ºC in air-lift PBR and after 40 days, we reached to the death phase. Normal time was about 15-21 days and in comparison with the normal one, it was too late and considerable (Figure 3).

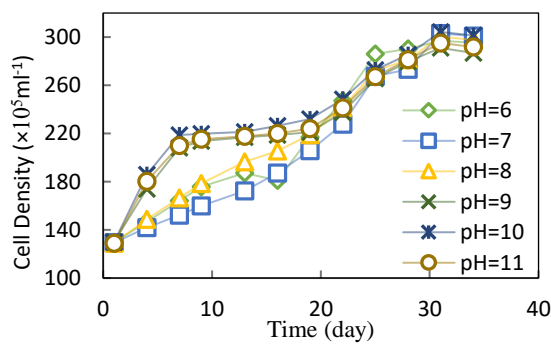

**Fig. 1** Growth at different pH

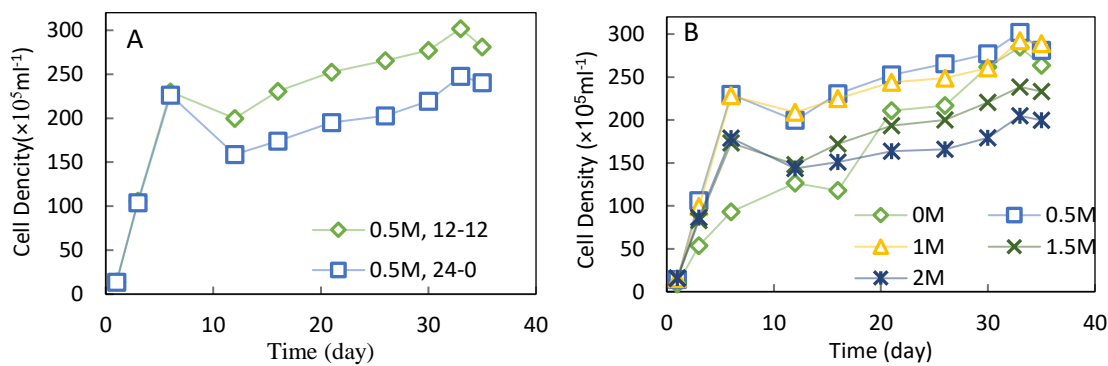

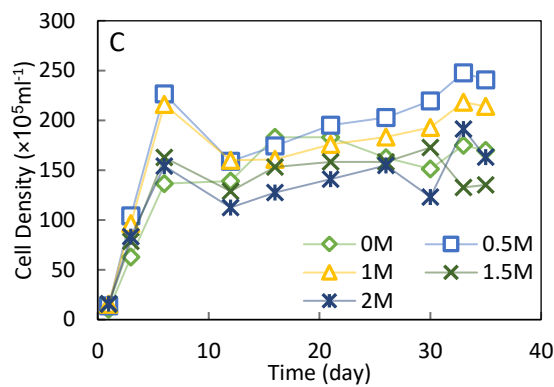

**Fig. 2** Growth at different daylight: A) in 0.5M, B) in 12-12 daylight and different concentration of NaCl, and C) in 24-0 daylight and different concentration of NaCl.

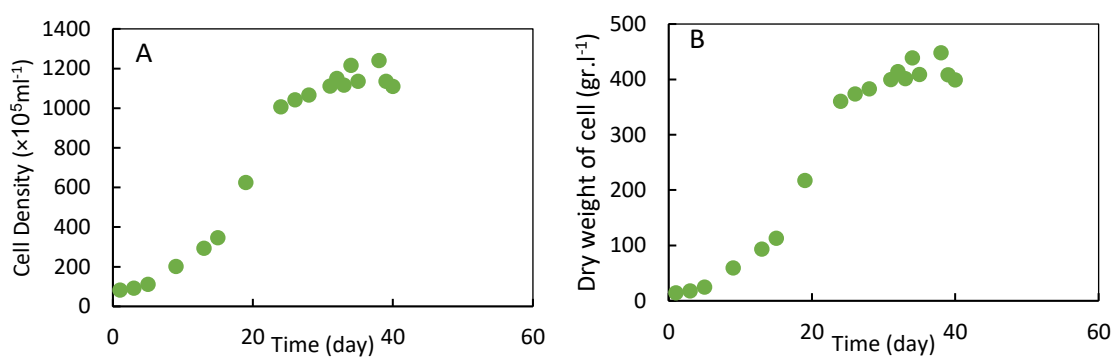

**Fig. 3** Growth chart: A) cell density and B) biomass

### 3.2. The Effect of Salt Shock in Growth and Pigment Concentration

At first day of death phase, we started salt shock entirely in different concentration. As can be seen in Figure 4, after salt shock, the growth of 0.5 M culture increased, while the growth of about 1M was milder and variation of cell densities was insignificant. About other cultures, they led to the reduction in growth. The growths of cultures in different concentrations of salt shock were not significantly different with each other and this indicates that salt shock doesn't have significant influence on growth as except in 0.5M.

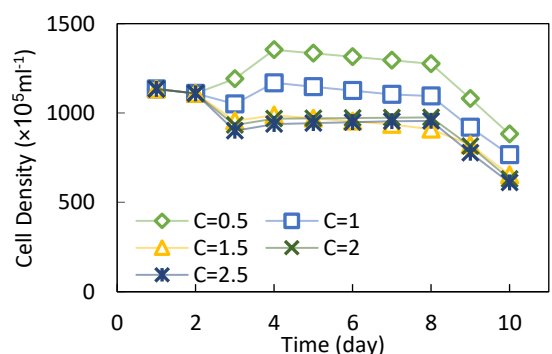

**Fig. 4** Cell density after salt shock

Chlorophyll *a* accumulation of all cultures rapidly increased after salt shock. The highest chlorophyll

*A* content appeared in 1.5 and 2M and the lowest one was observed in 2.5M (figure 5a).

Chlorophyll *b* accumulations of all cultures were so different with chlorophyll *A*. Figure 5b exhibits that at different concentrations, chlorophyll *B* accumulation experienced both decreasing and increasing trends after passage of time. However, it was seen that after 10 days, the 2.5M showed the best accumulation of chlorophyll *b* and the least chlorophyll B content appeared in 0.5M.

Accumulation of *β*-carotene in different trials was investigated after salt shock and the highest yield of *β*-carotene in this cultivation was 58.12 mg.l$^{-1}$. The result obtained through all trials shows 0.5M as the best concentration of salt shock for this pigment. After that, 2M yielded the highest accumulation (Figure 6).

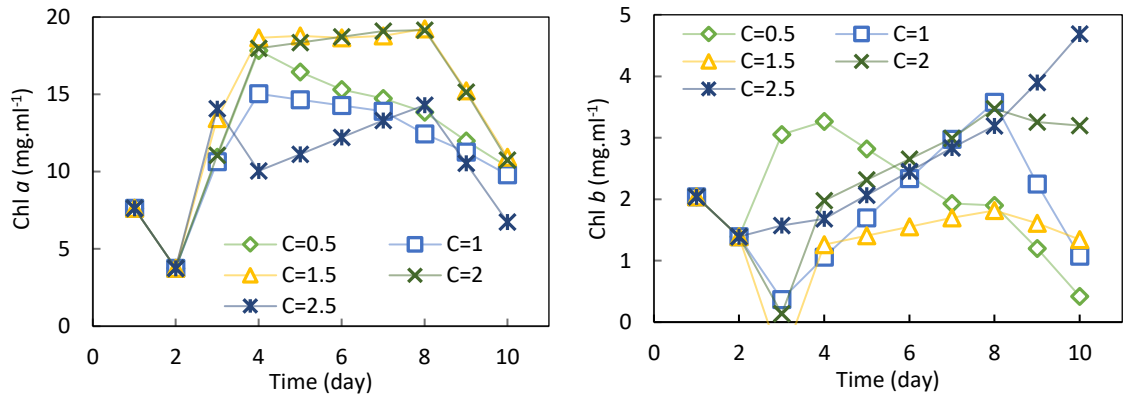

**Fig.5** Accumulation of Chl *a* and Chl *b* after salt shock

Total carotenoid accumulation had its highest value in 0.5M. For all cultures, we had some exotic changes in second day of shock. However, in 6$^{th}$ day of shock, we observed the highest content of carotenoeid and then rapidly decreased in all cultures (Figure 7).

According to the observations, during cultivation and after that we obtained 0.5M salt shock as best for growth and accumulation of most of pigment about this microalga. Also because of investigation of salt concentration and pH coincide, we found that Na ion had more positive effect on growth in comparison with Cl ion. This result obtained because we saw the best growth in pH=10 (with high Na) and it was more than pH=6 (with high Cl). There is an optimum point for amount of NaCl and about these microalgae this is 0.5M.

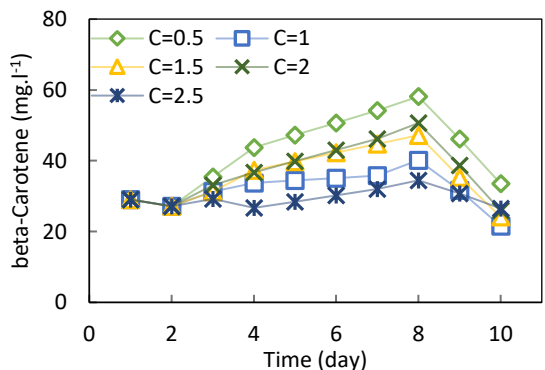

Fig. 6 Accumulation of beta-carotene after salt shock

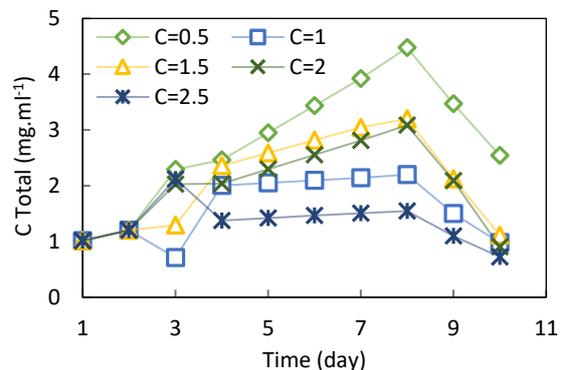

Fig. 7 Accumulation of total carotenoid after salt shock

## 4. Acknowledgments

The authors would like to thank Green Technology Laboratory (GTL) and also Sarvin Biotechnology Laboratory (SBL) for experimental support in faculty of chemical engineering at Tehran University.

## 5. References

[1] Ben-Amotz, A., Effect of irradiance and nutrient deficiency on the chemical composition of Dunaliella bardawil Ben-Amotz and Avron (Volvocales, Chlorophyta). Journal of plant physiology, 1987. 131(5): p. 479-487.

[2] Priyadarshani, I. and B. Rath, *Commercial and industrial applications of micro algae–A review.* J algal biomass utln, 2012. 3(4): p. 89-100.

[3] Ben-Amotz, A. and M. Avron, *The biotechnology of mass culturing Dunaliella for products of commercial interest.* Algal and cyanobacterial biotechnology, 1989: p. 91-114.

[4] de Jesus, S. and R. Maciel Filho, Modeling growth of microalgae Dunaliella salina under different nutritional conditions. Am J Biochem Biotechnol, 2010. 6: p. 279-283.

[5] Ben-Amotz, A., J.r.E. Polle, and D. Subba Rao, *The alga Dunaliella*. 2009: Science Publishers.

[6] Becker, E.W., *Microalgae: biotechnology and microbiology*. Vol. 10. 1994: Cambridge University Press.

[7] Çelekli, A. and G. Dönmez, Effect of pH, light intensity, salt and nitrogen concentrations on growth and β-carotene accumulation by a new isolate of Dunaliella sp. World Journal of Microbiology and Biotechnology, 2006. 22(2): p. 183.

[8] Prieto, A., J.P. Canavate, and M. García-González, Assessment of carotenoid production by Dunaliella salina in different culture systems and operation regimes. Journal of biotechnology, 2011. 151(2): p. 180-185.

[9] Ben-Amotz, A., New mode of Dunaliella biotechnology: two-phase growth for β-carotene production. Journal of applied phycology, 1995. 7(1): p. 65-68.

[10] Katz, A. and U. Pick, *Plasma membrane electron transport coupled to Na+ extrusion in the halotolerant alga Dunaliella.* Biochimica et Biophysica Acta (BBA)-Bioenergetics, 2001. 1504(2): p. 423-431.

[11] Tafreshi, A.H. and M. Shariati, Pilot culture of three strains of Dunaliella salina for β-carotene production in open ponds in the central region of Iran. World Journal of Microbiology and Biotechnology, 2006. 22(9): p. 1003-1006.

[12] Gomez, P.I., A. Barriga, A.S. Cifuentes, and M.A. Gonzalez, Effect of salinity on the quantity and quality of carotenoids accumulated by Dunaliella salina (strain CONC-007) and Dunaliella bardawil (strain ATCC 30861) Chlorophyta. Biological Research, 2003. 36(2): p. 185-192.

[13] Del Campo, J.A., M. García-González, and M.G. Guerrero, *Outdoor cultivation of microalgae for carotenoid production: current state and perspectives.* Applied microbiology and biotechnology, 2007. 74(6): p. 1163-1174.

[14] Ben-Amotz, A. and M. Avron, On the factors which determine massive β-carotene accumulation in the halotolerant alga Dunaliella bardawil. Plant Physiology, 1983. 72(3): p. 593-597.

[15] Sathasivam, R. and N. Juntawong, *Modified medium for enhanced growth of Dunaliella strains.* Int J Curr Sci, 2013. 5: p. 67-73.

[16] Lichtenthaler, H.K. and C. Buschmann, *Chlorophylls and carotenoids: Measurement and characterization by UV-VIS spectroscopy.* Current protocols in food analytical chemistry, 2001.

# Authors' background

| Your Name | Title | Research Field | Personal website |
|---|---|---|---|
| Mahsa Yazdani | | Biotechnology, Environment, Nanotechnology | - |
| Omid Tavakoli | Assistant Professor | Environment, Energy, Biotechnology, Biorefinery of microalgae | - |